\documentclass[11pt]{article}
\usepackage{array}
\usepackage{graphicx}
\usepackage{textpos}
\usepackage{geometry}   
\usepackage{dsfont}
\usepackage{amsthm}
\usepackage{amsmath}
\usepackage{amssymb}
\usepackage{esint}
\usepackage{mathrsfs}
\usepackage{graphicx}
\usepackage{amssymb}
\usepackage{epstopdf}
\usepackage{textpos} 

\numberwithin{figure}{section}

\title{Theory Vision, LHCP 2016\footnote{Invited presentation at LHCP Conference, Lund, June 2016. }}

\author{Frank Wilczek\\
\small\it Center for Theoretical Physics, MIT, Cambridge MA 02139 USA \\
\small\it Department of Physics, Stockholm University, Stockholm Sweden \\
\small\it Department of Physics and Origins Project, Arizona State University, Tempe AZ 25287 USA \\
\small\it Wilczek Quantum Center, Zhejiang University of Technology, Hangzhou 310023 China
}
\begin{document}

\maketitle

\begin{textblock*}{5cm}(11cm,-8.2cm)
\fbox{\footnotesize MIT-CTP-4836}
\end{textblock*}

\begin{abstract}
I give my perspective on promising directions for high energy physics in coming years. 
\end{abstract}

\medskip

\bigskip

\section{The Standard Model: Glory and Discontents}

1. Our standard model, or core theory, understood to consist of the $SU(3)\times SU(2) \times U(1)$ gauge theory of strong and electroweak interactions, together with minimally coupled Einstein gravity -- the gauge theory of local Lorentz invariance -- is a glorious achievement.  It provides a firm foundation for chemistry, astrophysics, and all practical forms of engineering.    It justifies and fulfills the great reductionist program of understanding the physical world based on precise mathematical understanding of the interactions among a few basic building-blocks. 

Consider, for example, the direct numerical solution of QCD.   Here there are no uncontrolled approximations, no perturbation theory, no cutoff, and no room for fudge-factors.   A handful of parameters, inserted into a highly constrained theory of extraordinary symmetry, either will or won't account for the incredible wealth of measured phenomena in the strong interaction.   And it certainly appears that they do.

\begin{figure}[h!]
\begin{center}
\includegraphics[scale=0.6]{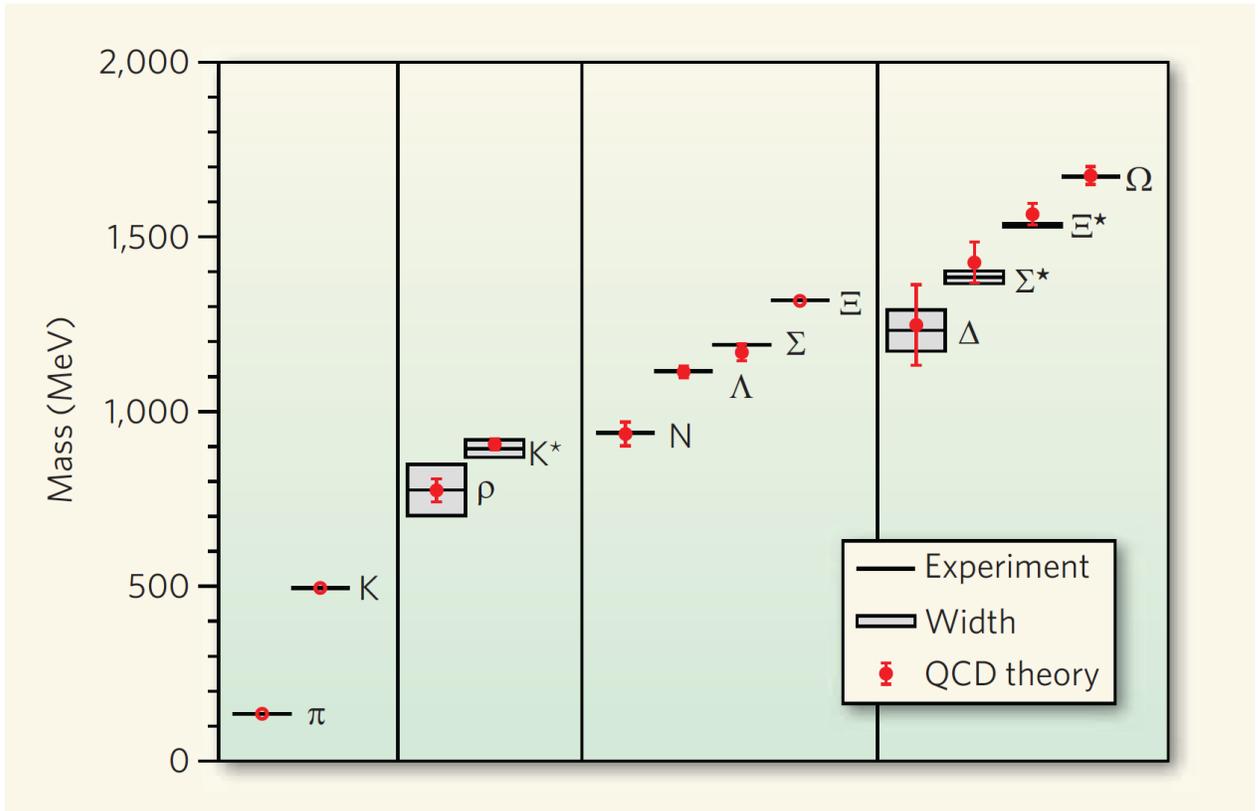}
\caption{Comparison of theory and experiment in the calculation of low-lying hadron masses.  Input parameters are the average $\frac{m_u + m_d}{2}$ of up and down quark masses, $m_s$ the strange quark mass, and the strong coupling constant, which determines the overall mass scale. \cite{qcdFigure}}
\label{qcdFigure}
\end{center}
\end{figure}

Among the rest, we see here the origin of (most) mass in pure energy, according to $m = E/c^2$: Using massless gluons and (almost) massless quarks as our ingredients, we account for the mass of protons and neutrons.  

If the LHC project does nothing other than confirm the standard model, it will have made a profound contribution to science and culture.   

2.  Yet that would be a disappointing result.  For all its virtues, the standard model does not have the look of a finished product.   Its symmetry structure is imperfect, with quarks and leptons  falling into several lopsided multiplets.  For unknown reasons, those multiplets are triplicated.   Then there is the Higgs doublet, which by contrast comes in singly, and whose many couplings, fixed to accommodate quark and lepton masses and mixings, are poorly constrained theoretically.  

There are also important phenomena, notably in cosmology, which the standard model does not address.  The astronomical dark matter, in particular, appears to be a gas of relic particles, but none of the standard model's particles has the right properties.  Also, we would like be more specific about promising but vague fundamental physics-based scenarios for the origin of matter-antimatter asymmetry and of inflation, and to get better insight into the nature of the dark energy (Einstein's cosmological constant).

3. There are, of course, many directions of theoretical exploration around the issues in fundamental physics that inspire the LHC program.  In this short ``visionary'' talk I will have to be very selective.   I will mainly focus on a pair of ideas that I think deserve to be true.  You may not agree with that assessment, but I hope you will agree that they are ideas whose truth-value it is important to determine.

\bigskip

\section{Unification}

\bigskip

\subsection*{Quantum Numbers}

4. The product gauge symmetry structure $SU(3)\times SU(2) \times U(1)$ practically begs to be embedded into a larger, encompassing symmetry.   The electroweak theory, which breaks $SU(2) \times U(1)_Y \rightarrow U(1)_Q$, shows how the full symmetry of fundamental equations can be hidden in their low-energy solutions, by the influence of cosmic fields or condensations (Higgs mechanism).   Slightly more elaborate versions of the same mechanism can implement
$SU(5) \rightarrow SU(3)\times SU(2)\times U(1)$ or $SO(10) \rightarrow SU(3)\times SU(2)\times U(1)$.   

An important test for the hypothetical expanded symmetries is whether they act naturally on quarks and leptons.   Indeed, another ``imperfection'' of the symmetry of the standard model is that it classifies quarks and leptons into several unrelated multiplets, even within one family.  If we allow for the right-handed neutrino $N$, needed to give a smooth theory of neutrino masses, there are six (if not, five).   Moreover the $U(1)_Y$ hypercharge quantum numbers we need to assign to those multiplets are funny fractions, determined phenomenologically.   

It is remarkable that the simplest candidate symmetries to unify the product groups of $SU(3)\times SU(2) \times U(1)$ into a (technically) simple structure -- $SU(5)$, and especially $SO(10)$ -- do a brilliant job of organizing the fermion multiplets and explaining those funny fractional hypercharges \cite{georgiGlashow}.

(Though I won't develop it here, I should mention that one can constrain the hypercharges in an alternative way, by demanding anomaly cancellation.   That approach does not address the unification of couplings, which I'll discuss momentarily.  Nor does it explain the multiplet structure nearly so neatly: In particular, it does not predict the existence of the right-handed neutrino $N$, which plays a central role in the theory of neutrino masses.)

\begin{figure}[h!]
\begin{center}
\includegraphics[scale=0.6]{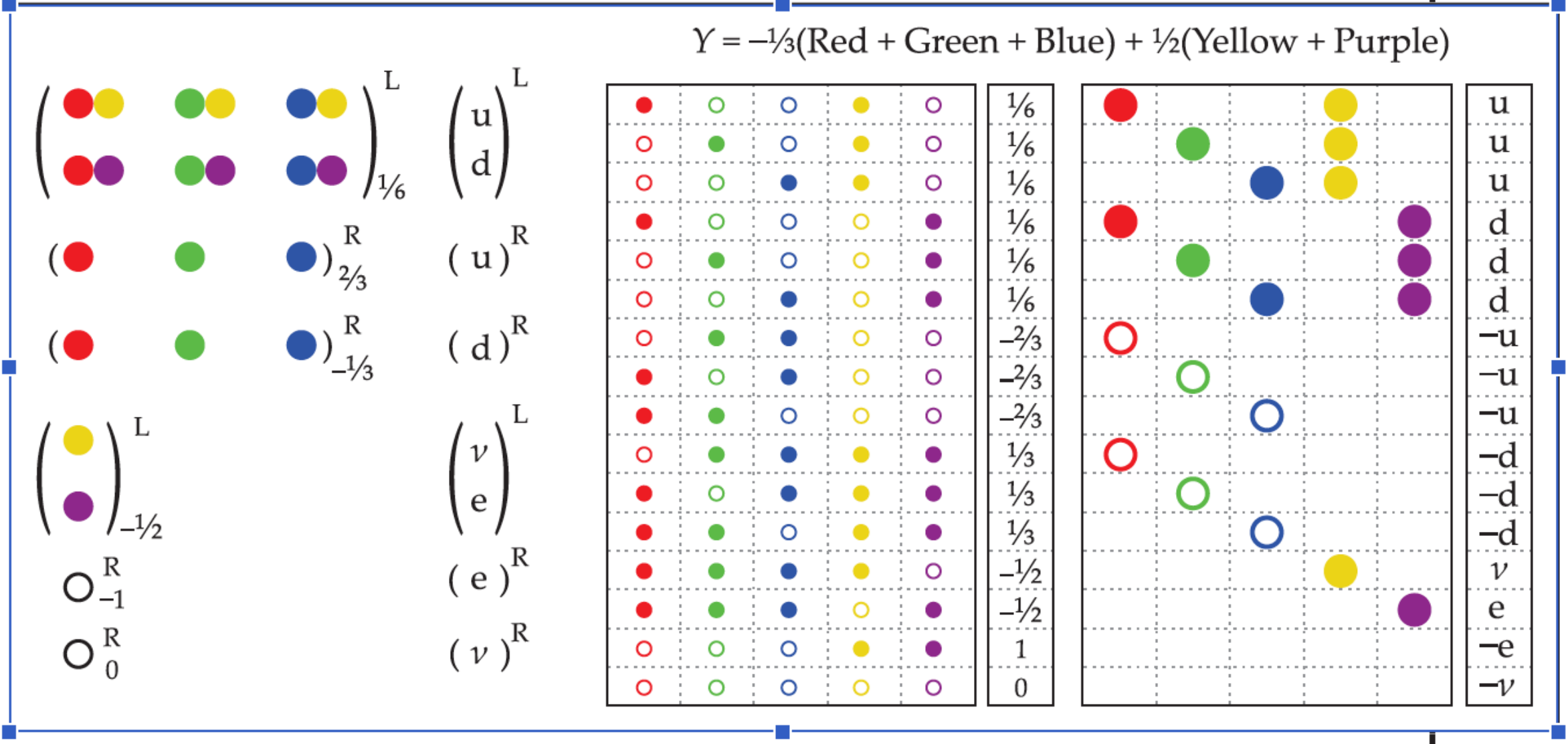}
\caption{This figure shows how the rather untidy organization of fundamental particles, according to our present day core theories of the strong, weak, and electromagnetic interactions falls neatly into, and can be explained by, a highly symmetric unified theory.}
\label{unificationFigure}
\end{center}
\end{figure}

I reviewed the details of quantum number unification recently elsewhere \cite{maxwell150}, and I will not repeat the mathematical analysis here.  Let us proceed directly to the summarizing Figure \ref{unificationFigure}.

The black column of symbols in the left-hand panel shows up (u) and down (d) quarks, the electron (e), and the electron neutrino ($\nu$).   Each of those particles can have a left-handed (L) or right-handed (R) helicity; the left-handed particles are organized in doublets, so the column displays six distinct entities.  The left side of the table displays the properties of those entities that account for their strong, weak, and electromagnetic interactions.  Gluons, which mediate the strong interaction, respond to three strong ``color'' charges, here indicated by red, green, and blue.  The weak interactions, which only act upon left-handed particles, respond to two weak color charges, here indicated as yellow and purple.  Electromagnetism is incorporated through couplings to electric charge.  On the left side of the panel, the average charge within each entity, also called its hypercharge ($Y$), is indicated by a numerical subscript.  

The right hand panel shows how the observed, scattered pattern of particles can be deduced from a unified template, derived from higher symmetry.  The far-right column of the panel again gives the names of the particles.  Here, everything has left-handed helicity; right-handed particles in the earlier description are now represented through their left-handed antiparticles, so that $u^R$ becomes $-u$ and so forth.  The far-left table shows all possible assignments of full or empty circles of strong and weak colors, subject to the constraint that the number of full circles is even.  A full circle is interpreted as a positive half-unit of the corresponding charge, an empty circle as a negative half-unit.  The hypercharge values, which appear in the middle column, are now generated from the weak and strong color charges according to the formula on top. 

We get to the right side of the table panel by invoking the rule that adding equal amounts of all the strong color charges, or equal amounts of all the weak color charges, corresponding to singlet quantum numbers, is invisible to the strong or weak interactions.  That allows us to change the color scheme on the left side of the table into that on the right side.  Note that the color charges are now in full units.  Remarkably, the resulting strong and weak colors, and the hypercharges, neatly match the quantum numbers of the entries the right-hand panel, i.e. reality.  (Note that antiparticles have the opposite color charges and hypercharges from their corresponding particles.)

\subsection*{Coupling Strengths}

5. The glory of local (gauge) symmetry, however, is that it controls not only bookkeeping, but also dynamics.   For a simple (in the technical sense) gauge group  such as $SU(5)$ or $SO(10)$, symmetry predicts all the couplings of the gauge bosons, up to a single overall coupling constant.    Thus unification predicts relationships among the strong, weak, and hypercharge couplings.  Basically -- up to the group-theoretic task of normalization --  it predicts that the three couplings for $SU(3)\times SU(2)\times U(1)$ must be equal.   
As observed, of course, they are not.  But the two great dynamical lessons of the standard model -- namely symmetry breaking through field condensation (Higgs mechanism), and running of couplings (asymptotic freedom) -- suggest a way out \cite{gqw}.  We can imagine that the symmetry breaking $G \rightarrow SU(3)\times SU(2) \times U(1)$ occurs through a big condensation, at a high mass scale.   In the symmetric theory, appropriate to the description of processes at large mass scales, there was only one unified coupling.   But we make our observations at a much lower mass scale.  To get to the unified coupling, we must evolve the observed couplings up to high energy, taking into account vacuum polarization.  Note that throughout that evolution the unified symmetry is violated, so the three $SU(3)\times SU(2) \times U(1)$ couplings evolve differently.   

Let us pause to consider, in broad terms, what we can expect from this sort of calculation.  Our input will be the observed couplings, plus some hypothesis $\cal H$ about the spectrum of virtual particles we need to include in the vacuum polarization.   Our output should be the unified coupling strength, and the scale of unification.   For any given $\cal H$, we have three inputs -- the observed couplings -- and two outputs -- the scale and coupling at unification.  So there will be a consistency condition.  If the calculation works, we will have reduced the number of free parameters in the core of the standard model by one, from three to two.   There are additional {\it physical\/} consistency conditions, concerning the value of the unification scale, which are quite significant.   I'll come to those momentarily.

\begin{figure}[h!]
\begin{center}
\includegraphics[scale=0.6]{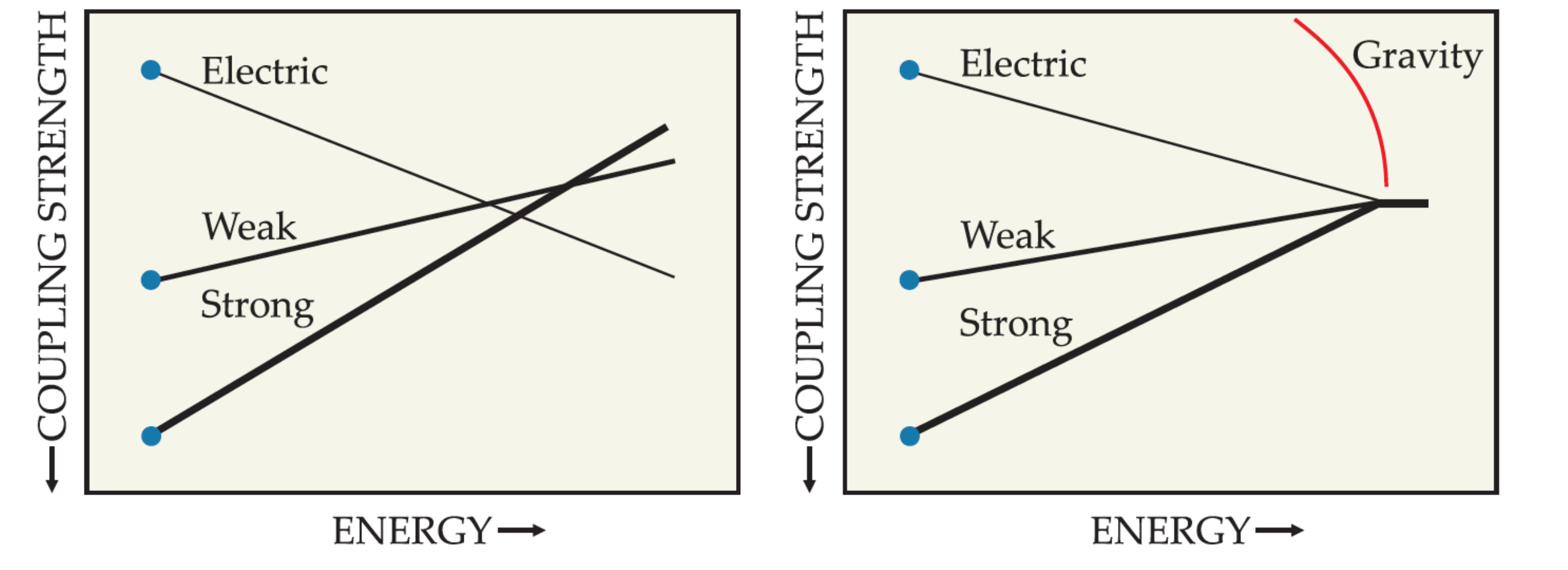}
\caption{The equality of coupling strengths at high energy, suggested by theories of unification, is suggested, but not achieved, by minimal extension of known interactions (left panel).   With the additional hypothesis of supersymmetry, one achieves good quantitative agreement.  For more detailed discussion, see \cite{maxwell150}.}
\label{couplingFigure}
\end{center}
\end{figure}

Again, I reviewed the details of coupling unification recently elsewhere \cite{maxwell150}, and I will not repeat the mathematical analysis here. Instead, let us proceed directly to the iconic summarizing Figure \ref{couplingFigure}.

The minimal hypothesis $\cal H_{\rm min.}$ is to take into account vacuum polarization from all the presently known particles, and no others.   Doing that, we find a suggestive approach to equality, but a quantitative failure, as illustrated in the panel on the left.

\subsection*{Supersymmetry}

6. There are several ways to motivate the additional hypothesis of supersymmetry.  Here I'll mention one that is rarely (if ever) discussed, but which I find profoundly attractive.   It is especially apt, in the context of unification.  

Wave-particle duality, a central achievement of quantum theory, allows us to treat light and ``material'', or ultimately force and substance, in a unified fashion.   The quantum-mechanical treatment of single photons or single electrons, for example, is essentially identical.  That unification breaks down, however, when we move beyond single particles, to consider assemblies.   There we find a division of the world into bosons and fermions, whose quantum statistical properties are diametrically opposed.   Supersymmetry, however, allows us to transform between those two statistical types, and thereby completes the work of force-substance unification.

7. Supersymmetry requires that we add new particles.  Let us suppose that we do this in a minimal way, adding as few particles as possible, and keeping their mass as low as possible.  With that hypothesis $\cal H_{\rm susy}$ of low-energy supersymmetry as input, our modified coupling strength unification \cite{drw} succeeds quantitatively.

The Planck energy
\begin{equation}
{\cal E}_{\rm Planck} ~=~ \sqrt{ \frac {\hbar c^5}{8 \pi G_N} } ~\approx~ 2.4 \times 10^{18} \, {\rm GeV}
\end{equation}
is another famous energy scale that can be constructed from fundamental constants\footnote{We have quoted the so-called rationalized Planck scale, including the factor $8\pi$ that naturally appears with $G_N$ in the Lagrangian of general relativity.}.  Here the construction is simple dimensional analysis, based on Newton's gravitational constant $G_N$ together with $\hbar, c$.    On the face of it, Planck units set the scale for effects of quantum gravity; thus when we consider basic (technically: ``hard'') processes whose typical energies are of order $E$, we expect gravitational effects of order $(E/{\cal E}_{\rm Planck})^2$.   

Our scale ${\cal E}_{\rm unification} \approx 2 \times 10^{16}$ GeV is significantly, but not grotesquely, smaller than the Planck scale.  This means that at the unification scale the strength of gravity, heuristically and semi-quantitatively, is of order 
\begin{equation}
({\cal E}_{\rm unification}/{\cal E}_{\rm Planck})^2 ~\sim~ 10^{-4}
\end{equation}
to be compared with the strength $g_5^{\, 2} /4\pi ~\sim~ 10^{-2}$ for the other interactions.   The relative smallness of gravity, thus estimated, which is of course further accentuated at lower energies, suggests that our neglect of quantum gravity in the preceding calculations may be justified.  

On the other hand, it seems to me remarkable that the comparison comes so close.  A classic challenge in fundamental physics is to understand the grotesque smallness of the observed force of gravity, compared to other interactions, as it operates between fundamental particles .   Famously, the gravitational interaction is $\sim 10^{42}$ times smaller than 
any of the other forces.   Again, however, proper comparison requires that we specify the energy scale at which the comparison is made.   Since the strength of gravity, in general relativity, depends on energy directly, it appears hugely enhanced when observed with high-energy probes.  At the scale of unification ${\cal E}_{\rm unified} \sim 2\times 10^{16} \, {\rm GeV}$ the discrepant factor $10^{42}$ is reduced to $\sim 10^4$, or even a bit less.  While this does not meet the challenge fully, it is a big step in the right direction.  

By expanding our theory, unification along the lines we have been discussing brings in additional interactions.  The two classic predictions for ``beyond the standard model'' interactions are small neutrino masses, leading to neutrino oscillations, and proton decay.   The first has been vindicated; the second not yet.   In both cases, the large scale ${\cal E}_{\rm unification}$ is crucial for explaining the smallness of the new effects.  For an authoritative review of these and other aspects of unification, emphasizing the phenomenological issues, with many further references, see \cite{raby}.

To me, the results of \S 5, supplemented by the discussion of \S 6 and \S 7, provide powerful, if circumstantial, {\it quantitative\/} evidence for the hypothesis of low-energy supersymmetry.   We've ``seen'' the effects of its particles, in their virtual form.   Now we look for LHC to start producing them as real particles.

8. Unfortunately, however, this circle of ideas does not yield a sharp estimate for the mass of the new particles.  The vacuum polarization effects, through which we have ``seen'' them, depend only logarithmically on their mass values.  While the best fits seem to indicate a few TeV mass scale for the gluinos, and less for the other gauginos, an order of magnitude heavier is not excluded; while the squarks and sleptons are very weakly constrained, and could be considerably heavier.   Though the LHC has a great opportunity to discover the new  particles, a negative result would not be definitive.   

\bigskip

\section{T Symmetry and Its Violation}

\bigskip

9. Few aspects of experience are as striking as the distinction between past and future.  Yet for the first three centuries of recognizably modern physics, ever since the Scientific Revolution, it appeared that the fundamental laws knew no such distinction.  They obeyed time-reversal symmetry, or simply T.   Newton's theories of dynamics and gravity, Einstein's modifications thereof, quantum electrodynamics, and all experiments in particle and nuclear physics appeared to display that feature.  As a result T was taken for granted, and came to be regarded as a fundamental principle in its own right.   

That situation changed in 1964, when a team led by James Cronin and Val Fitch, working at Brookhaven National Laboratory, discovered a subtle T-violating effect in the decay of K mesons (unstable particles produced at high-energy accelerators) \cite{croninFitch}. 
Their result presented a tremendous challenge to theoretical physics: To understand why T works extremely accurately in most circumstances, given that it is {\it not\/} a fundamental principle. 

Progress in fundamental physics over a broad front, culminating in the establishment of today's standard model, essentially in its modern form, greatly illuminated this issue.    Physicists realized that general principles of special relativity, quantum mechanics, and gauge symmetry together powerfully constrain the possible interactions of particles.  (Those general principles might eventually fail, of course, but they were key to formulating the standard model, and they have survived very rigorous testing.  Until proven otherwise, most physicists accept them as working hypotheses, more fundamental than T.)    When applied to the known build-blocks of matter -- quarks, leptons, gauge bosons, the graviton, and the Higgs particle -- those general principle determine that there are exactly two possible forms of T-violating interactions.  We might call them weak and strong T-violation, since they are mainly associated with the weak and strong interaction, respectively.       

10. Weak T-violation was elucidated in a brilliant paper by Makoto Kobayashi and Toshihide Maskawa, in 1973 \cite{km}.  Their theory relied on the existence of new particles, not then discovered: what we call today the bottom and top quarks $b, t$.   Those quarks were later discovered.   The Kobayashi-Maskawa theory explains the original Cronin-Fitch observation, and also many follow-up measurements of weak interaction processes.

11. Strong T symmetry remains problematic, however.

As I've emphasized, the interactions allowed by the standard model are constrained by powerful general principles of quantum mechanics, relativity, and gauge symmetry.  Given those constraints, every possible interaction has been measured to occur -- with one prominent exception.  The exception is a possible interaction among the color gluon fields, the so-called theta term, governed by the Lagrangian density
\begin{equation}\label{thetaTerm}
\Delta {\cal L} ~=~ \frac{g_s^2 \theta}{32\pi^2} \, {\tilde G}^{a \mu \nu} G^a_{\mu \nu}  ~=~ \frac{g_s^2 \theta}{8\pi^2} \, {\vec E}^a \cdot {\vec B}^a
\end{equation}
Here $g_s$ is the strong $SU(3)$ coupling constant, $a = 1, ... , 8$ is an index parameterizing the adjoint representation (octet of color gluons), the $G^a_{\mu \nu}$ are gluon field strengths and the $ {\tilde G}^{a \mu \nu}$ their duals.  In the second equality we re-write this expression in non-relativistic form, using the color electric and magnetic fields ${\vec E}^a , {\vec B}^a $.  $\theta$ is a dimensionless parameter.  

This term, Eqn.\,(\ref{thetaTerm}),  is manifestly local, relativistically invariant, and gauge invariant.  Furthermore its mass dimension is 4, and since QCD is asymptotically free, one expects that this interaction can be included consistently, without upsetting the good ultraviolet behavior of the theory, which is necessary to insure its existence. The new term opens up the possibility of strong T-violation.

Yet, as we shall soon discuss, all presently available observations are consistent with $\theta = 0$.  Existing experimental bounds correspond to $| \theta | \lesssim 10^{-10}$.  This is a mockery of dimensional analysis, and is often said to be ``unnatural''.

\subsection*{Electric Dipole Moments}

12.  At present the best bounds on $\theta$ arise from measurements of fundamental electric dipole moments.  The electric dipole moment of the electron appears as the coefficient of an interaction term (Hamiltonian)
\begin{equation}
H ~=~ - {\vec d}_e \cdot \vec E ~=~ - d_e \, \frac{\vec S}{| \vec S |}  \cdot \vec E
\end{equation}
governing the coupling of its spin to an electric field.   Here the subscript $e$ denotes ``electron''; there are of course similar definitions for other particles.  Since the electric field is a natural (polar), time-reversal even vector, while the spin is an unnatural (axial), time-reversal odd vector, the electric dipole moment interaction is odd under both spatial inversion and time reversal transformations.   Thus non-zero electric dipole moments reflect violation of the corresponding symmetries.    

(Note that the familiar electric dipole moments of elementary chemistry involve transition matrix elements.  They are appropriate to use when the energy separation between the states involved can be neglected, which is often the case in practical work.) 

There is a long history of attempts to measure fundamental electric dipole moments, going back to pioneering experiments by Purcell and Ramsey.     Among the most important present bounds are \cite{pospelov}
\begin{eqnarray}
| d_{\rm ^{205}Tl} | ~&<~  9 \times 10^{-25} \  {\rm e-cm} \nonumber \\
| d_{\rm ^{199}Hg} |  ~&<~  2 \times 10^{-28} \ {\rm e-cm} \nonumber \\
| d_{\rm n} | ~&<~ 6 \times 10^{-26} \ {\rm e-cm}
\end{eqnarray}
for atoms based on the indicated isotopes of tellurium and mercury, and neutrons.  

There are promising new ideas to greatly improve the accuracy of such measurements, specifically for the electric dipole moment of the proton \cite{protonEDM} using accelerator techniques similar to those that have been successful in giving precise measurements of the muon's magnetic dipole moment.  These are extremely important measurements, both the pin down the $\theta$ parameter and to probe other possible sources of T violations, such as arise abundantly in low-energy supersymmetry.

Since the size of the neutron, as measured either by its Compton wavelength or its charge radius, is in the neighborhood of $10^{-14}$ cm, and it contains several quarks whose charges are robust fractions of e, naive dimensional analysis would suggest a value more in the neighborhood of $10^{-15}$ e-cm.   The mismatch -- a suppression factor of $10^{-10}$ or less -- is impressive.  More precise reasoning and calculation \cite{pospelov} leads to the quantitative bound 
\begin{equation}\label{thetaBound}
| \theta | \lesssim 3 \times 10^{-10}
\end{equation}

\subsection*{Axions}

13. To ascribe the appearance of such an extraordinarily small number in the fundamental laws of nature to coincidence is a poor response.   It seems likely that Nature is trying to tell us something, and that we have been given an opportunity to expand our understanding of fundamentals.  

Several speculative proposals have been put forward to explain the smallness of $\theta$, but only one has been widely accepted.   Its basic idea is due to Roberto Peccei and Helen Quinn \cite{pq}.  It involves postulating a new kind of symmetry, called Peccei-Quinn (PQ) symmetry.   

Steven Weinberg and I, independently, discovered a key consequence of PQ symmetry, which its authors had overlooked \cite{wein} \cite{wil}.  We realized that it implies the existence of a qualitatively new kind of particle, the {\it axion}.   (I named the axion after a laundry detergent, noting that it cleaned up a problem with axial currents.)   The axion helps one visualize how PQ symmetry does its job: The axion field screens the potential T-violating interaction, analogously to how electrons in a conductor screen electric charge.  

Due to their very specific connection with symmetry, one can calculate the expected properties of axions in considerable detail, given the value of one parameter $F$, the scale of symmetry breaking, which the theory does not determine.   They are predicted to be extremely light spin-0 particles, whose interactions with ordinary matter are extremely feeble.   As $F$ gets larger both the mass and the interactions grow smaller, in proportion.   For $F = 10^{12}$ GeV, the axion mass is about $10^{-5}$ eV.  There have been many attempts to discern signals of axions, including work at accelerators and astronomical observations.   Taken together, they indicate $F \ge 10^{10}$ GeV.  

John Preskill, Mark Wise and I discovered a most remarkable cosmological implication of axions \cite{axionCosmo}.  If one includes the axion field into the evolution of matter through the big bang, one finds that a very substantial density of axions survives, initially in the form of a cold Bose-Einstein condensate, which later gets stirred and mixed by gravitational self-attraction.   Indeed, if $F \ge 10^{11}$ GeV, the axion fluid becomes an excellent candidate to supply a significant, and possibly the dominant, contribution to the mysterious dark matter of the universe.  Its density is estimated to be sufficient, and its other properties are consistent with the observed properties of dark matter.   

An enormous literature has grown up around axion physics \cite{axionReview}, and several international conferences have been either wholly or in large part devoted to the subject.   While the central ideas have evolved and matured, they have survived many years of intense scrutiny.  On the other hand, no other comparably attractive approach to the strong P, T problem has  emerged.   

%(They include important aspects I will not discuss here, notably including the laboratory experiments and solar observations that constrain axions in the approximate range $10^9 \, {\rm GeV} \, \le F \le 10^{10} \, GeV$.)  

It has become a most important goal, both for fundamental physics and cosmology, either to observe the cosmic axion fluid or to rule it out.  This is a widely shared perception, as evidenced by a recent spike in activity on this subject around the globe. 

The most mature approach to cosmic axion background detection relies on a strategy first proposed by Pierre Sikivie \cite{sikivie}.   This strategy exploits the basic interaction of axion electrodynamics, which allows an axion, in the presence of a magnetic field, to convert into a photon.   A determined, sophisticated program, named ADMX (Axion Dark Matter eXperiment) \cite{admx}, centered at the University of Washington, Seattle, is pioneering this approach experimentally.  The new CAPP (Center for Axion and Precision Physics) initiative in South Korea \cite{capp} proposes to develop magnet technology that will enable a next-generation experiment.   Very recently Huaixiu Zheng, Matti Silveri, R. T. Brierley, S. M. Girvin and K. W. Lehnert \cite{girvin} have brought ideas from cavity QED into the discussion, which promises to accelerate the search process.    The Sikivie strategy is best suited to $10^{11} \, {\rm GeV} \,  \le F \le 10^{13} \, {\rm GeV}$, but within that range, given this ferment of activity and ideas, it promises to reach the level of sensitivity required to detect the cosmic axion background.   

Newer ideas, which may allow access to larger $F$ values, are presently in early stages of development.   The CASPEr (Cosmic Axion Spin Precession Experiment) strategy relies on ability of a cosmic axion background to induce a small oscillating electric dipole moment in nuclei \cite{casper}.  In the presence of perpendicular magnetic and electric fields, this will cause precession of the nuclear spins, which in turn induces a tiny but potentially detectable oscillating magnetic field.  The ABRACADABRA (A Broadband/Resonant Approach to Cosmic Axion  Detection with an Amplifying B-field Ring Apparatus) strategy is again based on the fundamental interaction of axion electrodynamics, but uses a different geometry, better adapted to low frequencies, whereby in the presence of a magnetic field the axion background induces a potentially observable oscillating magnetic field \cite{abra}. 

\bigskip

\section{Is That All There Is?}

\bigskip

\subsection*{Other Interactions}
 
14. When the talk was given, the fate of the 750 GeV gamma-gamma resonance was undecided.  I said that it seemed gratuitous, and that its existence would be giving us an unwanted lesson in humility.   The preceding suggestions for unification among all the forces, and for achieving fundamental understanding of T violation, seem to me compelling -- as I've said, they deserve to be true -- and I'd hate to have to walk them back.   The resonance would have been difficult to explain without new superstrong interactions, or even more radical departures from the ideas that underlie quantitative coupling unification.  It is hard to see how those departures could fail to spoil that success.   

Of course Nature gets the last word, but we get to set our priors, and mine made that resonance seem improbable.   

15. On the other hand new $SU(3) \times SU(2) \times U(1)$ singlet sectors generally do not interfere with unification or axion ideas, and their existence is not implausible.   Scalar fields, such as the Higgs field, by providing low-dimension effective interactions, may open portals to those sectors. 

\subsection*{Other Worlds}

16. John von Neumann concluded his thoughtful address ``The Mathematician'' \cite{vN} with the following warning:
\begin{quote}
As a mathematical discipline travels far from its empirical source, or still more, if it is a second and third generation only indirectly inspired by ideas coming from ``reality'' it is beset with very grave dangers. It becomes more and more purely aestheticizing, more and more purely l'art pour l'art. ... whenever this stage is reached, the only remedy seems to me to be the rejuvenating return to the source: the re-injection of more or less directly empirical ideas. I am convinced that this was a necessary condition to conserve the freshness and the vitality of the subject and that this will remain equally true in the future.
\end{quote}
His warning applies, I think, even more forcefully to theoretical physics.   

In our quest to understand fundamental processes, we've built up a powerful world-describing machinery.  We've learned that symmetry and topology, in the context of quantum physics, are extremely powerful conceptual forces, shaping the behavior of matter.    We've learned that Nature, embodied in quantum field theory, is a reliable source of interchangeable parts.  And we've learned that ``empty space'' or ``vacuum'' is best conceived as a dynamical medium, both richly responsive (vacuum polarization) and substantial (condensates).    These lessons can be applied creatively to craft materials which, viewed from the inside, are worlds with novel properties which, when accessed from the outside, are both interesting and useful.   Omar Khayyam's aspiration 
\begin{verse}
Ah, Love! could thou and I with Fate conspire \\
To grasp this sorry Scheme of Things entire! \\
Would not we shatter it to bits-and then \\
Re-mould it nearer to the Heart's Desire!
\end{verse}
can be our inspiration.

\subsection*{Thinking Way Ahead}

17. Stars are powered by nuclear energy, and in a few tens of billions of years they will run down.   That will create a very constraining environment for our descendants' activities.   Fortunately, there may be a way out.   Magnetic monopoles are a generic feature of unified theories, and they can catalyze a further stage of burning, by allowing  exothermic reactions such as 
\begin{equation}
{\rm Nucleus} (A, Z) + e ~\rightarrow~ {\rm Nucleus} (A-1, Z-1) \, + \, {\rm energy}
\end{equation}
to proceed rapidly \cite{rubakov} \cite{callan}.   Such reactions release far more energy than conventional nuclear interactions.  So one can imagine a post-nuclear world, whose energy-based economy continues for an additional several hundred billions of years.   

We are left with the challenge of producing the monopoles.  If their mass is in the expected $10^{18}$ GeV range, that will not be easy.   We will need major advances in accelerator technology, and a resolute building program.   Given the stakes, it is not too soon to start investing in the R+D, and to continue to build prototypes with ever-increasing energy reach.

\bigskip

\bigskip

{\it Acknowledgement}: FWÕs work is supported by the
U.S. Department of Energy under grant Contract Number DESC0012567 and by the Swedish Research Council under grant Contract No. 335-2014-7424.

\end{document}